\documentclass[draft,grl]{agu2001}
\usepackage{lineno}
\usepackage{graphicx}
\authorrunninghead{V\"{O}R\"{O}S ET AL.}

\titlerunninghead{MAGNETIC FLUCTUATIONS AND TURBULENCE}

\linenumbers*[1]

\begin{document}

%
%
%
%
%

%
%

\title{Magnetic Fluctuations and Turbulence in the Venus Magnetosheath and Wake}
%

%
%


 \authors{Z. V\"or\"os, \altaffilmark{1}
 T. L. Zhang, \altaffilmark{2}
 M. P. Leubner, \altaffilmark{1}
 M. Volwerk, \altaffilmark{2}
 M. Delva, \altaffilmark{2}
 W. Baumjohann, \altaffilmark{2}
 and K. Kudela, \altaffilmark{3}}

 \altaffiltext{1}
 {Institute of Astro- and Particle Physics, University of Innsbruck, Innsbruck, Austria.}
 \altaffiltext{2}{Space Research Institute, Austrian Academy of Sciences, Graz, Austria.}
%
%
 \altaffiltext{3}{Institute of Experimental Physics, Slovakia Academy of Sciences, Kosice, Slovak Republik.}



%
%


\begin{abstract}
Recent research has shown that distinct physical regions in the Venusian induced magnetosphere are recognizable from the variations of strength and of wave/fluctuation activity of the magnetic field. In this paper the statistical properties of magnetic fluctuations are investigated in the Venusian magnetosheath, terminator,
and wake regions. The latter two regions were not visited by previous missions. We found 1/f fluctuations in the magnetosheath, large-scale structures near the terminator and more developed turbulence further downstream in the wake. Location independent short-tailed non-Gaussian statistics was observed.
\end{abstract}

%
%

%

\begin{article}

%
%
\begin{figure}
\noindent\includegraphics[width=30pc]{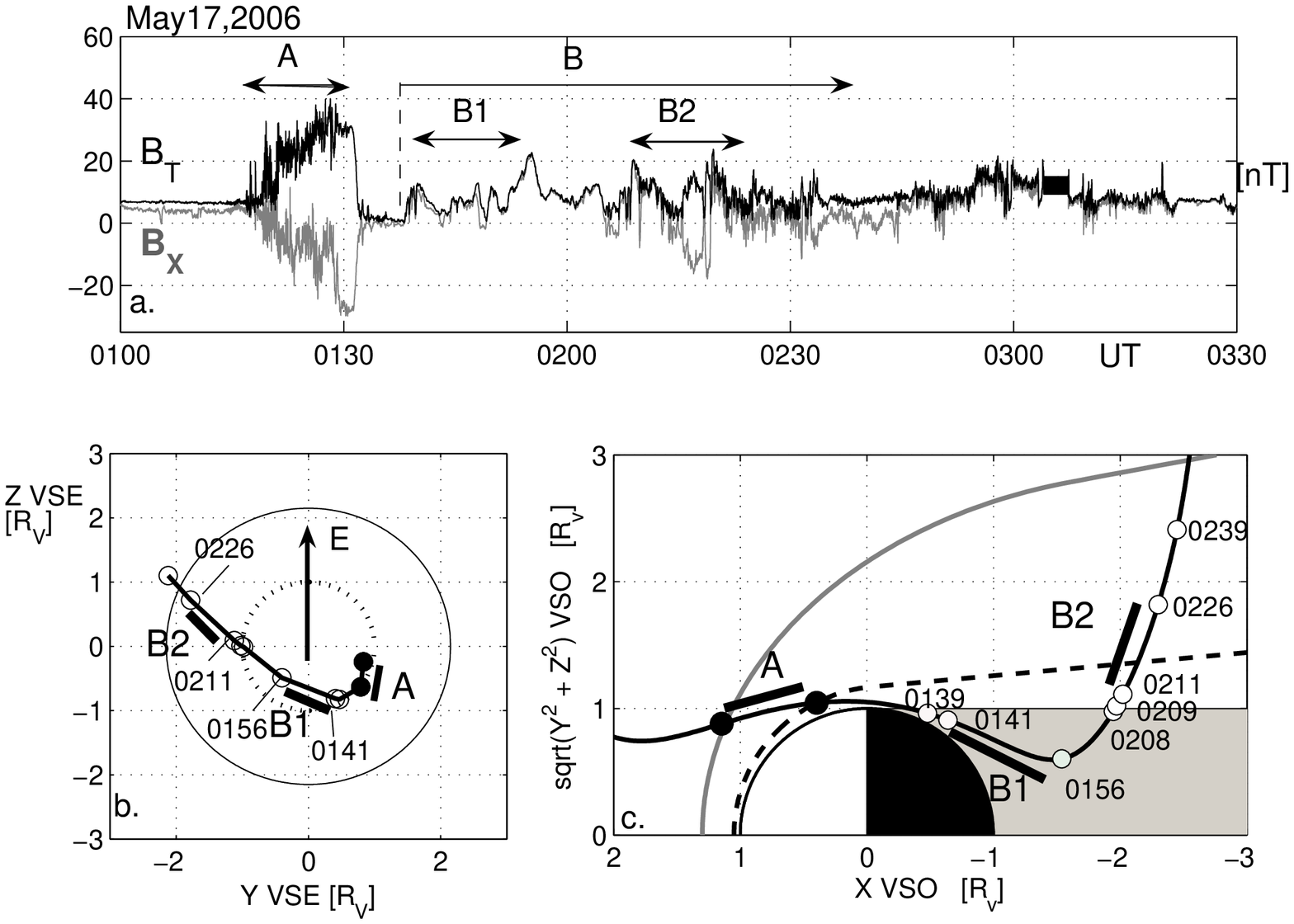}
 \caption{(a) Magnetic field strength $B_T$ (black line) and the $B_X$ component (grey line) on May 17, 2006; the capital letters A, B1 and B2 indicate intervals of equal lengths (black lines in each subplot) within the magnetosheath (A) and wake (B);  (b) VEX orbit (black line) in the VSE coordinate system of the convection electric field $E$;  time ticks on the orbit indicate the analyzed intervals; the optical shadow is indicated by the dashed circle, the continuous circle shows the terminator bow shock location;
 (c) VEX trajectory (solid black line) in VSO coordinate system; bow shock (solid gray line); induced magnetopause (dashed line); optical shadow (shaded region); time ticks on the orbit correspond to the analyzed intervals.}
\end{figure}

\begin{figure}
\noindent\includegraphics[width=40pc]{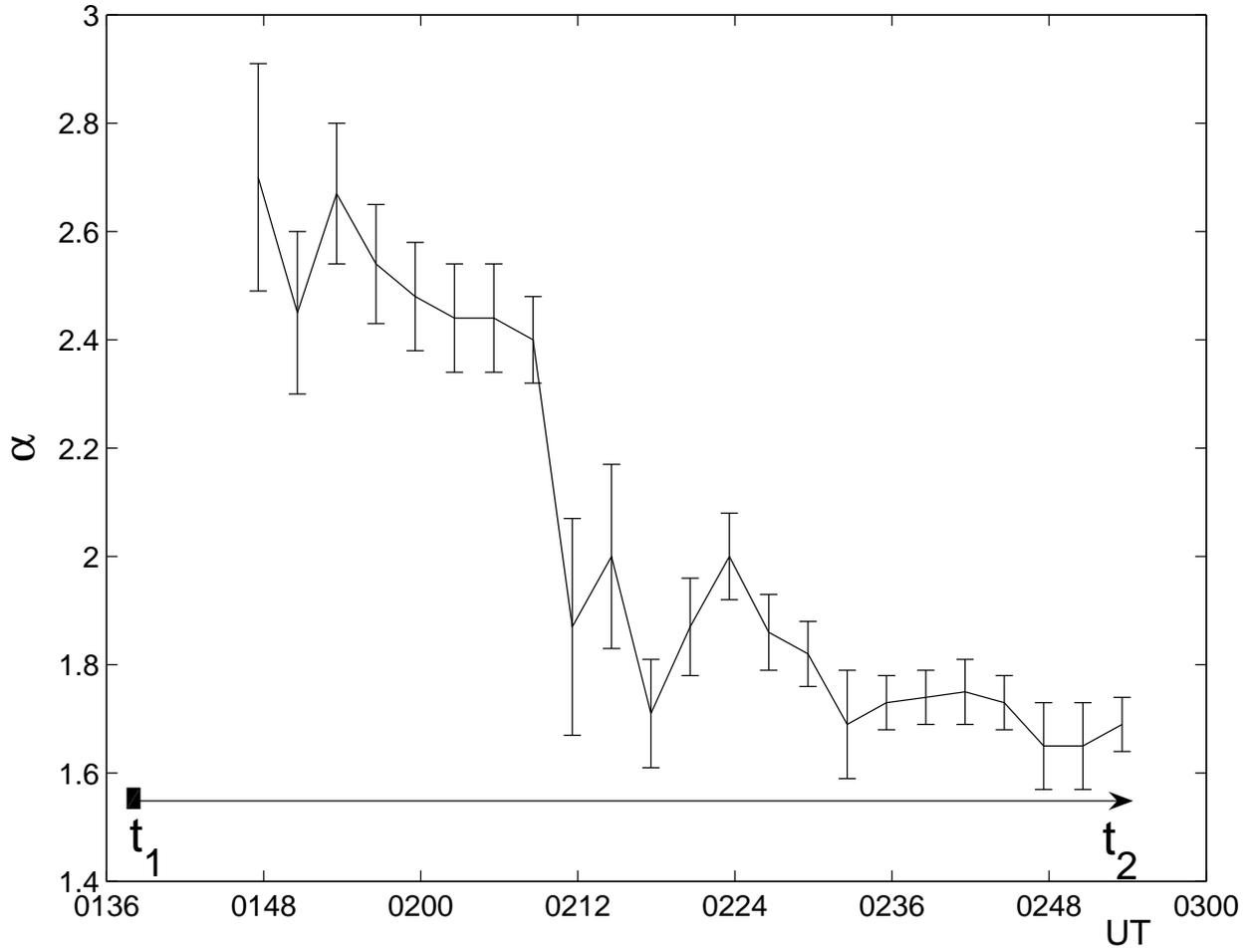}
 \caption{The change of spectral scaling index in the wake (interval B in Figure 1); $t_1$ = 0138 UT fixed, $t_2$ changed. }
\end{figure}

\begin{figure}
\noindent\includegraphics[width=40pc]{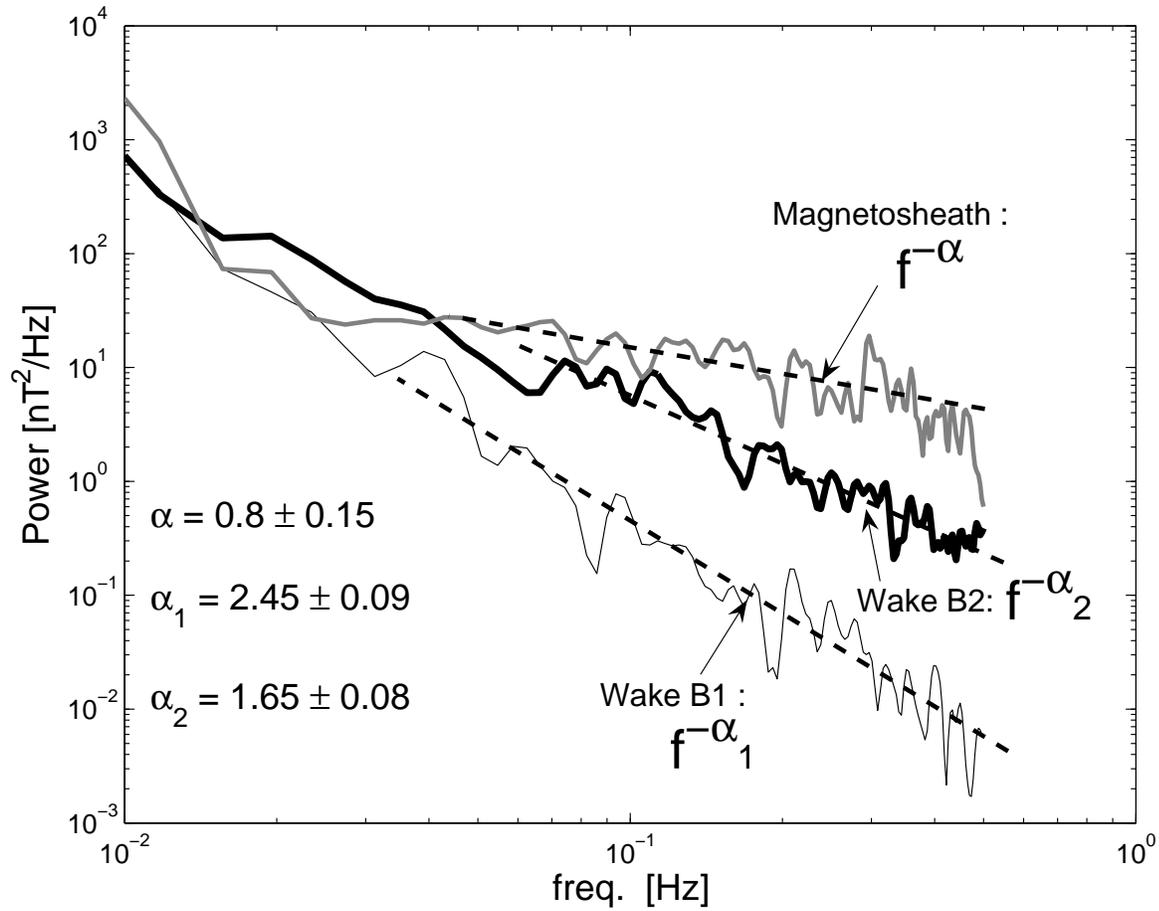}
 \caption{Power spectral densities for intervals of equal length in the magnetosheath (A), and in the wake (B1 and B2); the numerical values of spectral indices were obtained during data intervals with the smallest errors.}
\end{figure}

\begin{figure}
\noindent\includegraphics[width=40pc]{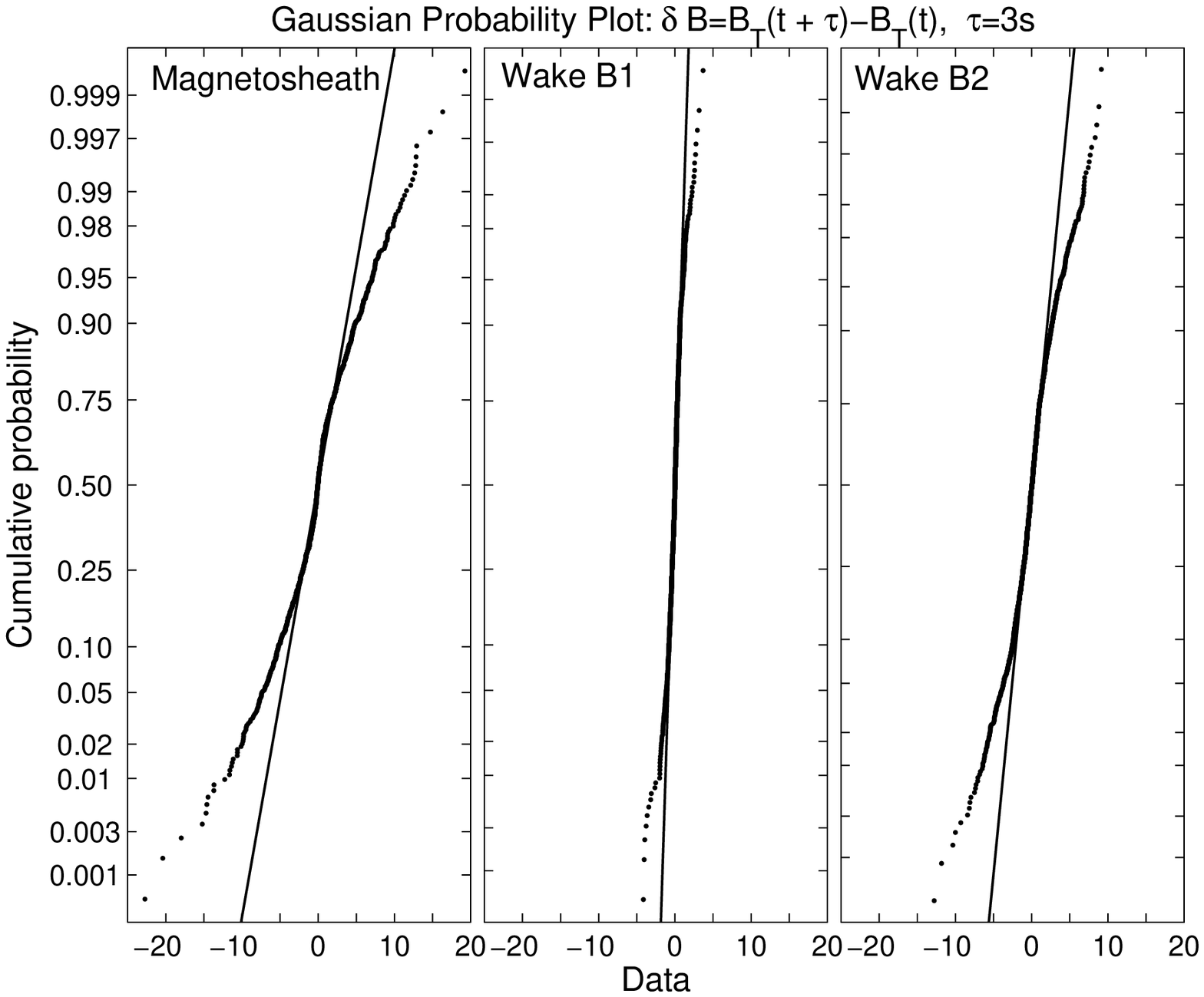}
 \caption{Gaussian (normal) probability plots for data intervals in the magnetosheath and wake. The solid lines correspond to the theoretical Gaussian, points correspond to the experimental data.}
\end{figure}

\section{Introduction}
In the absence of an intrinsic magnetic field, the structured plasma environment around Venus
is formed due to the direct interaction between the solar wind and the ionosphere. The ionosphere acts
as an obstacle to the supersonic solar wind flow carrying a magnetic field.
The interplanetary magnetic field
is draped around the planetary obstacle to form an induced magnetosphere (Zhang et al., 2007) which consists of the magnetic barrier on the dayside
and the magnetotail. The characteristic scales of the induced magnetosphere and the magnetotail are different from the spatial dimensions in the Earth's magnetosphere.
The results of Venus Express (VEX) provide a distance of the Venusian subsolar bow shock from
the surface of the planet of about 1900 km, while the terminator bow shock location is
2.14 $R_V$ ($R_V$ = 6051 km) at solar minimum.  The altitude of the Venusian induced
magnetopause is 300 km at the subsolar point and 1013 km at the terminator (Zhang et al., 2007).
For comparison, the width of
Earth's magnetosheath is about 25000 km at the subsolar point. Nevertheless, particularly in case
of quasi-parallel shock geometries, waves and fluctuations occur equally in both
Earth's and Venus' magnetosheaths (Luhmann et al., 1983, 1986). Due to the rather limited width of the
Venusian magnetosheath, however, the Reynolds number characterizing turbulent flows is presumably small.
As a consequence, the inertial scaling range (if it exists), corresponding to the scales with prevailing
nonlinear multi-scale interactions in turbulence, is narrow. Keeping in mind
that both the proton inertial length and the proton Larmor radius are of the order of 100 km, the magnetohydrodynamic (MHD) spatial scales
in the Venusian magnetosheath are limited to one or two decades in wave-number space.
For example, the Larmor radius is about 160 km for a proton temperature of 1 keV and B=20 nT.
The wider scaling region corresponds to the terminator shock location where the distance between
boundaries increases. The terminator and, further downstream, the night-side region is of particular interest,
where plasma instabilities, vortices and
wake turbulence can develop due to the planetary obstacle. The near-polar orbit of VEX
with a periapsis altitude of 250-350 km  allows for the first time observations at terminator and mid-magnetotail regions (Zhang et al., 2006) and thus to compare fluctuation statistics within
both magnetosheath and wake regions. These two important regions were not
covered by the previous missions, e.g. Pioneer Venus Orbiter (Russell, 1992).
Using 1-s resolution magnetic field data from Venus Express during one crossing we provide evidence of changing fluctuation/turbulence statistical characteristics along the VEX trajectory from the dayside magnetosheath to the wake (Figure 1c).

\section{Events on May 17, 2006}
The magnetic field strength $B_T$  (black line) and the $B_X$ component (grey line) on May 17, 2006 is shown in Figure 1a. Cartoons of the VEX trajectory (thick black lines) are shown in VSE
(Figure 1b) and VSO (Figure 1c) coordinate systems, respectively. In VSE, the X coordinate is towards the Sun, Z is in the plane of convection electric field $E = V_{SW} x B$, ($V_{SW}$ is the solar wind speed and $B$ is the IMF), and Y completes the orthogonal coordinate system. The plane of the plasma sheet is aligned with $E$ (e.g. Barabash et al., 2007).
VEX is crossing the bow shock (thick gray line, quasi-parallel shock in this case),
the magnetosheath, the magnetopause (dashed black line),
and the wake region (approximately the optical shadow - shaded region, Figure 1c). The events on this day were initially analyzed by Zhang et al. (2007), who demonstrated that a crossing of each
physical region is recognizable from the variations of strength and variation of the wave/fluctuation activity of the magnetic field.
The magnetic field strength is strongly fluctuating and its value increases up to $\sim$ 40 nT within the
magnetosheath (interval A in Figure 1a), after which $B_T$ decreases almost to zero (closest approach,
VEX is below the induced magnetopause). Magnetic field strength and fluctuations increase again during interval B
(Figures 1a, b) and this region is defined as the wake. Unlike the magnetosheath spatial extent, the wake location
is more difficult to determine. This is because, further downstream, VEX is again crossing
the magnetopause (at $\sim$ -2 $R_V$ in
Figure 1c), and returns to the magnetosheath and the solar wind (after 03:30 UT in Figure 1a).
Moreover, the induced magnetopause boundary can be more unstable downstream.
A statistical study of the induced magnetopause boundary crossings (Zhang et al., 2007) shows that there is a
significant uncertainty in the
boundary location between the terminator and X = -2 $(R_V)$. This indicates that VEX can be
located below or over the flapping
magnetopause in an unpredictable manner even during one crossing. In what follows we refer, for simplicity, to
fluctuations downstream of the terminator location, but before re-entering the solar wind,
as wake fluctuations/turbulence.

\subsection{Spectral scalings}

We determined the occurrence times of wake fluctuations through spectral analysis.
The spectral scaling indices $\alpha$, corresponding to magnetic field strength $B_T$, were estimated using a wavelet
method (V\"or\"os et al., 2004) for data intervals of increasing lengths $t_2 - t_1$, starting at $t_1$ = 01:38 UT (closest approach time). Figure 2 shows that, for $t_2$ extending to 0254 UT, two distinct scaling regions
appear, exhibiting a sudden transition at about 0208 UT.
The robustness of scaling index determinations were further examined for two sub-intervals before and after
the sudden transition at 0208 UT. Optimized data intervals with smallest errors
in $\alpha$ were found. Following this procedure, we identified two distinct time intervals
with well defined $\alpha$; one between 0139 and 0208 UT and the second between 0209 and 0239 UT. The corresponding time ticks on the orbit are shown in Figure 1c.
The corresponding values of spectral indices during these intervals are $\alpha_1 = 2.45 \pm 0.09$
and $\alpha_2 = 1.65 \pm 0.08$, respectively (Figure 3).
The spectral index in the magnetosheath, $\alpha = 0.8 \pm 0.15$, was obtained similarly,
for a shorter interval between 0116 and 0131 UT (filled circles in Figures 1b,c). The power spectral density plots in Figure 3
were computed from the whole magnetosheath interval ($A$) and two subintervals in the wake
($B_1$ and $B_2$, see Figure 1a), both having the length of $A$.
The time intervals $A$, $B_1$ and $B_2$ are
also indicated by thick black lines along the orbit in Figures 1b,c.
$B_1$ and $B_2$ are taken from the two longer, optimized wake intervals, for which the spectral
indices $\alpha_1$ and $\alpha_2$ were found previously.
The corresponding time ticks ($B_1$: 0141 - 0156 UT and $B_2$: 0211 - 0226 UT)   are
shown on the orbit in Figures 1b,c.

Figure 3 shows that over the frequency range of 0.04 - 0.5 Hz magnetosheath fluctuations (thick gray line)
exhibit larger power than wake fluctuations. The spectral index $\alpha$ is close to 1, which indicates
the presence of $1/f$ noise. We note  that the small peaks near 0.1 Hz might be associated with
mirror mode structures, which occur in planetary magnetosheaths due to temperature anisotropies (Volwerk et al., 2008, GRL, in press).
The methods and the data intervals  in this paper, however, were not designed for detection of
wavy structures, but for spectral scalings.

The wake fluctuations closer to the terminator (interval $B_1$, thin black line in Figure 3)
posses significant power only at
low frequencies, then the power rapidly decreases following $\sim f^{-2.45}$ (approximately as the
thick dashed line shows). Due to the shortness of the time series the peaks over the frequencies 0.04 - 0.1 Hz
are not statistically significant, but assuming bulk plasma velocity of 200 km/s near
the terminator (Spreiter and Stahara, 1992),
the largest power is contained in structures of 2000 - 5000 km size.

Numerical simulations indicate that the Kelvin-Helmholtz instability can occur at the terminator ionopause of Venus (Terada et al., 2002), capable of producing structures over 1000 km  and turbulence (Amerstorfer et al.,2007; Biernat et al., 2007). In fact, the initial VEX observation detected these waves (Balikhin et al., 2008, submitted to GRL). Our event study shows
that MHD turbulence near terminator is not developed because of the rapid decrease of spectral power toward
higher frequences. Further downstream, after 0209 UT (Figure 1) the spectral scaling index is
$\alpha_2 = 1.65 \pm 0.08$ (thick black line in Figure 3), very close to 5/3 or 3/2, which are the scaling index
values expected for hydrodynamic or MHD inertial range turbulence. Our spectral analysis shows that the
transition between regions with indices $\alpha_1$ and $\alpha_2$ is almost immediate, around 02:09 UT
(see also the waveforms in Figure 1a). We can speculate, therefore, that the spacecraft suddenly enters into a
wake region with more developed turbulence rather than gradually developing from the large-scale fluctuations observed near the terminator. The occurrence of turbulence during the interval $B_2$ might also be related to crossings of the plasma sheet. In fact, during this interval $B_X$ is changing sign several times and $B_T$ is decreasing locally (Figure 1a), which could be interpreted as a signature of multiple plasma sheet crossings. Figures 1b,c show, however,
that the interval $B_2$ is outside of the central part of the wake, that is, outside of the region where the $E$-field aligned plasma sheet could be situated  at $|Y VSE| < 0.5 R_V$ (e.g. Barabash et al., 2007).

\subsection{Comparison to Gaussian distributions}
Higher order statistics is needed to fully describe the nature of nonlinear fluctuations. In turbulent, boundary influenced non-homogeneous plasma flows, the shape of the probability density functions (PDFs) is scale-dependent, non-symmetric and peaked with long tails (e.g. V\"or\"os et al., 2007a). Because of the shortness of our time series the shapes of PDFs or their scale dependency cannot be evaluated. Instead, we construct Gaussian (normal) probability plots (GPPs) to decide whether or not our data are Gaussianly distributed. GPPs are plots of cumulative probabilities versus sorted data values with distorted y axis labels.  The latter ensures a linear dependency for the theoretical Gaussian distributions in GPPs (e.g. D'Agostino et al., 1990). We reconstructed GPPs using the magnetic field strength $B_T$ as well as two-point differences defined through $\delta B = B_T(t+\tau)-B_T(t)$, for $\tau = 2...30$ s. All these data sets taken from the magnetosheath and wake regions exhibit non-Gaussian statistics. A representative behavior for $\tau = 3$ s is depicted in Figure 4.
Points are experimental data while the solid lines correspond to theoretical Gaussian distribution. Near-terminator wake data (interval B1 in Figure 1a) are the closest to a Gaussian distribution. Nevertheless, all data sets exhibit nonlinearity, mainly the first few (negative data values, above the Gaussian line) and the last few (below the Gaussian line) experimental points. This is a characteristic behavior for non-Gaussian PDFs with short tails. Long-tailed data would deviate from the linear relationship in just the opposite way: the first few points would appear below
and the last few points above the Gaussian solid line (e.g. D'Agostino et al., 1990). The absence of long-tailed PDFs in case
of the turbulent wake (interval B2, Figure 1a) with scaling index $\alpha_2 = 1.65 \pm 0.08$ is surprising, because the statistics of developed turbulence is usually described in terms of long-tailed PDFs. The tails of PDFs are
built-up from rare but more energetic events, therefore, the observed short tails can just indicate that our time series are short and not representative enough to observe the expected long-tailed PDF. This assumption can be checked by investigation of similar crossings and multiple representations of wake turbulence near Venus.

\section{Conclusions}
In this paper the unique data from the VEX spacecraft were used to compare magnetic fluctuation statistics from  Venusian magnetosheath, terminator and mid-magnetotail regions. For simplicity, we denote the latter two regions  Venusian wake, never visited during previous missions.
To identify spectral scaling ranges and indices, we used a wavelet technique, successfully applied for studying the continuous spectra in the Earth's plasma sheet turbulence (V\"or\"os et al., 2004). Our data intervals with the smallest errors in spectral scaling index corresponded to different physical regions in real space.

We found 1/f non-Gaussian noise in the magnetosheath as signature of the presence of independent driving sources (see e.g. V\"or\"os et al., 2007b). The quasi parallel bow shock source for mirror mode waves, the instabilities and gradients near the magnetic field draping region - can probably contribute to the dayside fluctuations in the magnetosheath.

Near the terminator and further downstream in the Venusian wake, large-scale non-Gaussian structures are present, which can be partially explained in terms of the Kelvin-Helmholtz instability. As VEX continued further downstream into the wake,
more developed turbulence was detected with spectral scaling index typical for hydrodynamic or MHD turbulence in the inertial range of scales. The observed non-Gaussian short-tailed distributions can be caused by the shortness of our data sets. This calls for studying PDFs from multiple crossings and realizations of turbulent processes.

%
%

\begin{acknowledgments}
The work of Z.V. and M.P.L was supported by the Austrian Wissenschaftsfonds under grant number
P20131-N16 The work of K.K. was supported by the Slovak Research and Development
Agency under the contract  No. APVV-51-053805.
\end{acknowledgments}

\end{article}

\end{document}